\documentclass{article}

\usepackage{graphicx}          
\usepackage{dcolumn}          
\usepackage{bm}                  
\usepackage{float}                

\title{Solar-motion correction in early extragalactic astrophysics}

\author{Domingos Soares\footnote{\small dsoares@fisica.ufmg.br}~ and   
Luiz Paulo R. Vaz\footnote{\small lpv@fisica.ufmg.br} \\ Departamento de F\'{\i}sica, 
ICEx, UFMG --- C.P. 702 \\ 30123-970,  Belo Horizonte --- Brazil} 

\date{October 09, 2014}

\begin{document}

\def\Ho{$H_\circ$}
\def\omegao{$\Omega_\circ$}
\def\to{$t_\circ$}

\maketitle

\begin{abstract}
Redshift observations of galaxies outside the Local Group are fairly common in extragalactic 
astrophysics. If redshifts are interpreted as arising from radial velocities, these must be corrected 
by the contamination of the solar motion. We discuss the details of such correction in the way it was 
performed by the American astronomer Edwin Hubble in his 1929 seminal paper. The investigations   
of spiral nebulae undertaken by the Swedish astronomer Knut Lundmark, in 1924, are also considered 
in this context.     
\end{abstract}

\bigskip
\bigskip

\section{Introduction}
The light of distant galaxies --- at least outside the Local Group --- presents to the observer, specially 
to the spectroscopist, a singular feature discovered by the American astronomer Vesto 
Slipher (1875-1969) in the early decades of the XX century: the position of the spectral lines of the 
chemical elements are, in the great majority of cases, systematically displaced to wavelengths 
larger than those measured for the same elements (at rest) in earthly laboratories. Being the shifts 
in the direction of larger wavelengths, they are called \emph{redshifts} because, in the visible solar 
spectrum, red has the larger wavelength. Such a nomenclature is adopted even when the object's 
spectrum is outside the visible range.  Less frequently, there is also the \emph{blueshift}, whose 
definition is analogous to the redshift. 

What is the cause of the redshifts of galaxies?  Strictly speaking, this is still a question in dispute. 
One can, however, adopt the most obvious hypothesis, namely, that they originate from the motion 
of the galaxies, specifically due to a motion of recession from the observer. That is, the physical 
phenomenon responsible for the observed spectral shifts is the so-called \emph{Doppler effect}. 
Redshifts are usually represented by the letter z. For z$<$0.1, the recession velocity is then given 
by  v$\cong$cz, where c is the speed of light in vacuum  (Soares 2009, Fig. 4). 

Observations are done from the Earth, which has a rotational motion, that originates days and nights, 
and an orbital translation about the Sun.  These velocities are variable, depending on the time of 
the observation but they have amplitudes of 0.5 km/s and 30 km/s, respectively, which are, in general, 
much smaller than galaxy velocities. Even so, galaxy velocity observations are corrected for these 
motions and become \emph{heliocentric} velocities, that is, referred to the Sun.

One must, next, consider the motion of the Sun. This consists of the motion inside the Milky Way plus 
the motion of the Milky Way with respect to the general field of galaxies, or, as the American 
astronomer Edwin Hubble (1936, p. 106) prefers \emph{``with respect to the nebulae"}.  The 
observations of galaxies are expressed, as we saw, ``with respect to the Sun", and in order to have 
the motion of the galaxies with respect to the general field of galaxies one might remove the 
motion of the Sun with respect to this same field. We shall describe this procedure, in what follows,   
according to the method quantitatively prescribed by Hubble in his influential  article of 1929 and, 
rather clearly described in a qualitative way, in his book of 1936 entitled \emph{The Realm 
of Nebulae}.  In section 3, we apply Hubble's method to the galaxy sample of Knut Lundmark, Hubble's 
contemporary at astronomy.  We conclude with general remarks in section 4. 

\section{Solar motion correction}
In his book \emph{The Realm of the Nebulae},  Hubble explains how the motion of the Sun influences 
the motion of the ``nebulae" (i.e., of the galaxies; see Hubble 1936, p. 106): 
\begin{quotation}
\emph{Each observed velocity was thus a combination of (a) the ``peculiar motion” of the nebula, as the individual motion is called, and (b) the reflection of the solar motion (a combination of the motion of the sun within the stellar system [{\rm i.e., inside the Milky Way galaxy}] and the motion of the stellar system with respect to the nebulae). If sufficient nebulae were observed, their random peculiar motions would tend to cancel out, leaving only the reflection of the solar motion to emerge from the totality of the data.}
 
(\ldots) 

 \emph{Actually, the residual motions were still large and predominantly positive. The unsymmetrical distribution indicated the presence of some systematic effect in addition to the motion of the sun [{\rm with respect to the nebulae}] . }
\end{quotation}

(Texts in square brackets are ours).

The above-mentioned ``systematic effect" was modeled by Hubble simply as Kr --- a constant 
times the distance to the galaxy ---, differently of others in his days who added quadratic terms 
and even logarithmic ones in r.  

\begin{figure}[H]
\begin{center}
\includegraphics[width=14cm]{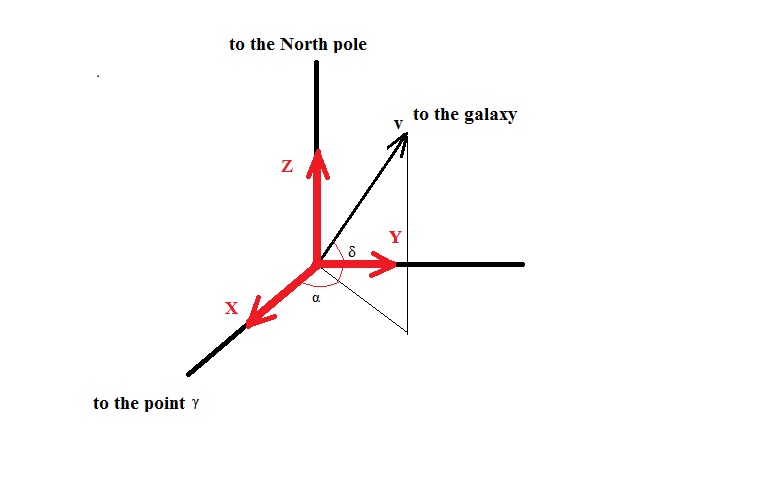}
\end{center}
\caption{The components X, Y and Z of the velocity symmetrical to the Sun velocity,  and the radial 
velocity v of a galaxy are displayed in the equatorial coordinate system. The galaxy right ascension is 
$\alpha$ and its declination is $\delta$. }
\label{fig:eqtc}
\end{figure}

Quantitatively, we follow Hubble (1929). The velocity of a galaxy observed from Earth, after 
the heliocentric correction, may be written, according to classical relativity, as 
the composition of two velocities: 
\begin{equation}
\label{eq:e1} 
\vec{v} \equiv \vec{V}_{G\odot} = \vec{V}_{GR} + \vec{V}_{R\odot} ~.
\end{equation}
The letter ``R" represents the ``reference frame of the nebulae". The motion of the reference 
frame of the nebulae with respect to the Sun  is given by the ``reflection of the solar motion" --- 
as Hubble writes --- with respect to the reference frame of the nebulae, that is: 
\begin{equation}
\label{eq:e2} 
\vec{V}_{R\odot} = (X, Y, Z) = -\vec{V}_{\odot R} = (-X_\odot,-Y_\odot,-Z_\odot) ~.
\end{equation}
On page 170 of Hubble (1929), the letter ``v" represents the \emph{radial} velocity of a galaxy 
measured with respect to the Sun, in other words, it is one of the components of the velocity 
vector.  According to the explanation given by Hubble above, we can then write the expression for 
v, wherein the ``systematic effect" proportional to the distance and the ``reflection of the solar 
motion" appear separately:  
\begin{equation}
\label{eq:e3} 
v = (\vec{v})_{radial}= Kr + (\vec{V}_{R\odot})_{radial} ~.
\end{equation}
Figure \ref{fig:eqtc} shows, in the equatorial system of coordinates, the velocity $(X, Y, Z)=(-X_\odot, 
-Y_\odot, -Z_\odot)$ and the velocity v of a given galaxy. The coordinates $\alpha$ and 
$\delta$ in the figure are, respectively, the right ascension and the declination of the galaxy. 
The projection of $(X, Y, Z)$ on the line of sight of a given galaxy --- the second term in the right-hand 
side of eq. \ref{eq:e3} ---can be derived from Fig. \ref{fig:eqtc} and is explicitly shown below: 
\begin{equation}
\label{eq:e4} 
v = Kr + X \cos \alpha\cos \delta + Y {\rm sen}~  \alpha\cos \delta + Z {\rm sen}~ \delta ~. 
\end{equation}
This is the very same equation that appears on page 170 of Hubble (1929). Through it and the 
galaxy observations (velocities and distances) we can get the solar motion with respect to the 
reference frame of the galaxies: 
\begin{equation}
\label{eq:e5} 
\vec{V}_{\odot R} = (X_\odot,Y_\odot,Z_\odot) = (-X,-Y,-Z) ~.
\end{equation}
We have in this problem four unknowns to be determined, K, X, Y and Z. In general, one has much 
more than four observed galaxies, and the resulting system of equations turns out to be 
overdetermined (more equations than the number of unknowns). This is a rather common situation 
in astrophysics. For example, a binary stellar system can be spectroscopically observed in many 
orbital phases yielding a set of observations much larger than the number of unknowns of the 
problem (orbital inclination and eccentricity, mass ratio, etc.). 
 
Next, we shall undertake such a procedure with the list of galaxies studied by whom is, 
by many regarded, one of the precursors of Hubble. In the end of the procedure we shall obtain 
the solar motion and the equivalent to the modern ``Hubble's constant" for the expansion of the 
galaxies.   

\section{The spiral nebulae of Knut Lundmark}
In 1924, the Swedish astronomer Knut Lundmark (1889-1958) published an article where he intended 
to determine the radius of curvature of the space-time, in the light of the cosmological model put 
forward by the Dutch physicist Willem de Sitter (1872-1934) in 1917. De Sitter's model predicted that 
light from a distant object should exhibit a redshift proportional to the object's distance. The radius 
of curvature would be determined from the constant of proportionality. Usually, redshift was interpreted 
as originating from the recession velocity of the object, calculated through the Doppler effect formula 
v=cz. Lundmark, hence, discuss diagrams velocity $\times$ distance for various classes of objects. 
We shall analyze his data for the so-called ``spiral nebulae", the modern spiral galaxies. 

\begin{figure}[H]
\begin{center}
\includegraphics[width=12cm]{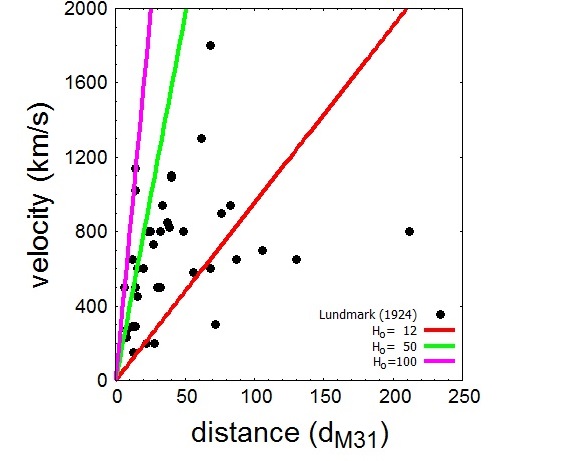}
\end{center}
\caption{The galaxies of Lundmark (1924, Table III) without solar motion correction. Distances to 
the galaxies are given in units of $\rm d_{\small M31}$, the distance to M31. Lines represent 
``Hubble's laws" with different Hubble's parameters, in (km/s)/Mpc, adopting the modern value of 
$\rm d_{\small M31}=784$ kpc. The slope $\rm H_\circ=12$ (km/s)/Mpc corresponds to the fitting 
of $\rm v=H_\circ d$ to Lundmark's data (see also Soares 2013). The accepted value of Hubble's 
constant nowadays is 72 (km/s)/Mpc with an uncertainty of 10\% (cf. Freedman et al. 2001 and 
Soares 2009).    }  
\label{fig:lund}
\end{figure}

Figure \ref{fig:lund} shows the data of Lundmark (1924, Table III). He determined the 
distances to the nebulae by comparing their apparent sizes and brightnesses with the size and 
brightness of M31. Thus distances are given in terms of the distance to M31 $d_{M31}$ (see 
more details in Soares 2013).  Adopting the modern value of $d_{M31}=$ 784 kpc (Stanek and  
Garnavich 1998), we superimpose on the data  some relations of proportionality of the form 
v=H$_\circ$d, where  H$_\circ$ represents the modern concept of ``Hubble's constant".  
The value H$_\circ$=12 (km/s)/Mpc is the slope of the line fitted to Lundmark's data, but 
forced to cross the origin. A linear fit to the data has a positive interception with the velocity 
axis of approximately 600 km/s and a slope of H$_\circ$=3 (km/s)/Mpc. The interception with 
the velocity axis  would indicate the contamination of the data by the solar motion. One can try then 
to remove the solar motion using the same proceeding adopted by Hubble in his work of 1929. 
In order to remove the solar motion we must solve an overdetermined system of equations, as 
seen in the end of section 2. We have a model (with a set of N unknowns or parameters --- N=4 
in our case, K, X, Y, Z) which must reproduce a series of M observations, being M much larger than 
N. It is this last feature that makes the system of equations to be called {\it overdetermined}. 

Our problem consists in making minimum the differences between the prediction of the model (the 
right-hand side of eq. \ref{eq:e4}) and the measurement of the observable (the radial velocity of a 
given galaxy, i.e., the left-hand side of eq. \ref{eq:e4}). In practice, what we minimize is the sum of 
the squared differences between the prediction (which we may call {\it calculated value}) and the 
observed value. This is the classical {\it method of least squares}.  

\begin{figure}[H]
\begin{center}
\includegraphics[width=6cm]{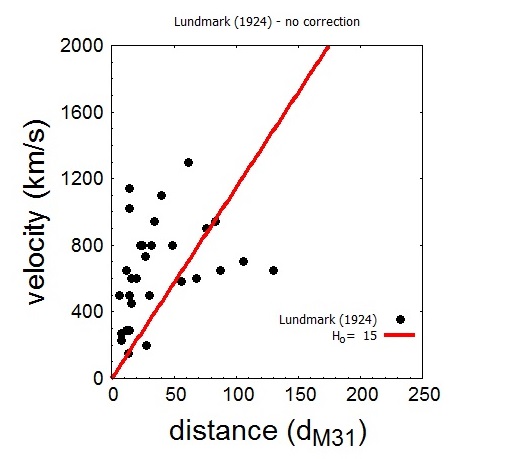}
\includegraphics[width=6cm]{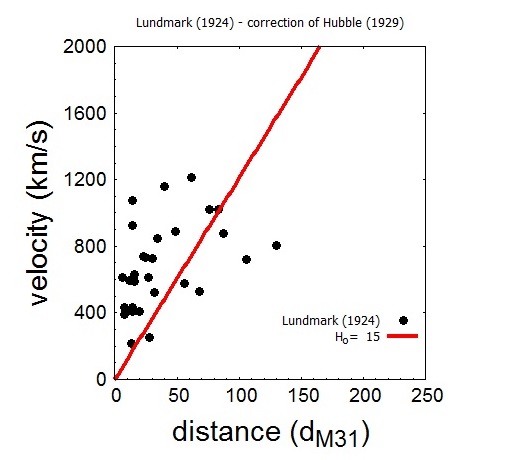}

\includegraphics[width=6cm]{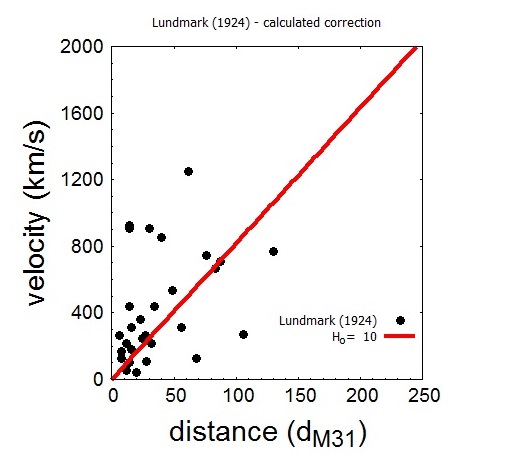}
\end{center}
\caption{The galaxies of Lundmark (1924), without galaxies with negative velocities (6) and without  
elliptical (6) and irregular (2) galaxies. From the 44 originally in Table III, 30 remain. The lines 
represent ``Hubble's laws" with different  $\rm H_\circ$ parameters in units of (km/s)/Mpc, 
adopting the modern value of   $\rm d_{\small M31}=784$ kpc.  Top-left panel shows galaxies 
without solar motion correction, top-right panel corrected by Hubble's 1929 solar motion and 
bottom panel with the calculated solar motion correction.  }
\label{fig:all}
\end{figure}

The results we obtained, by reproducing Hubble's 1929 methodology, implies into corrections 
in the observed velocities which, surprisingly, does not significantly affect the determination of the 
constant K (the modern ``Hubble's constant"). It is approximately the same either if we use 
the initial sample or the corrected one. To illustrate this aspect, we show below, in Fig. \ref{fig:all}, 
the diagrams V$\times$R for the sample of Lundmark (1924), without the negative velocities 
and, furthermore, without the elliptical and irregular galaxies, according to our present 
knowledge. Such galaxies are not appropriate to Lundmark's method of distance calculation, 
which is based in the comparison of apparent size and  brightness with a spiral galaxy, M31. 
This sample has 30 galaxies.   

\begin{figure}[H]
\begin{center}
\includegraphics[width=6cm]{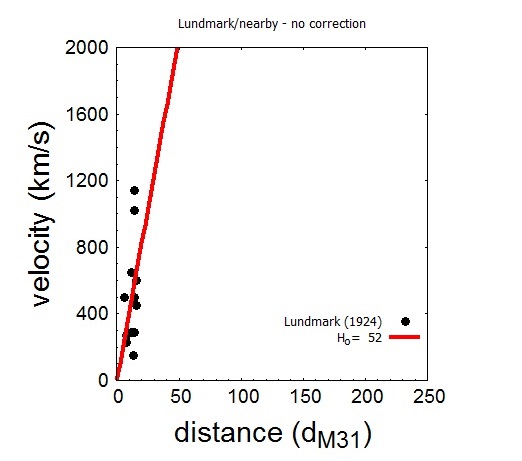}
\includegraphics[width=6cm]{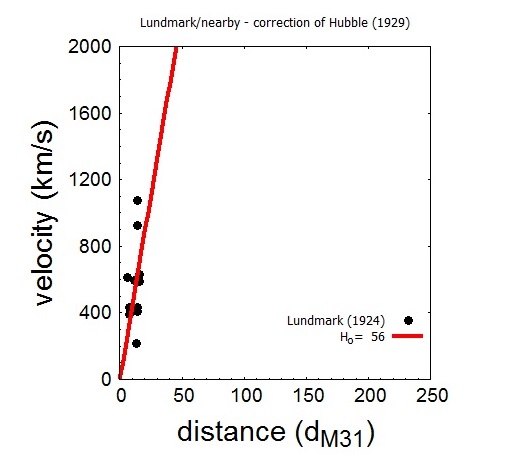}

\includegraphics[width=6cm]{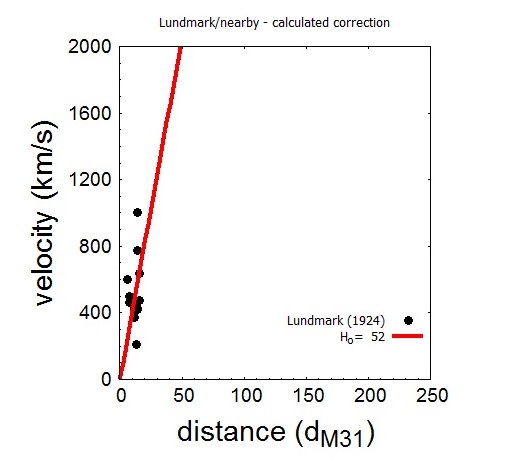}
\end{center}
\caption{The 12 nearest galaxies of Lundmark (1924), with distances consistent with Hubble's 
1929 sample. The lines represent ``Hubble's laws" with different  $\rm H_\circ$ parameters in units of (km/s)/Mpc, adopting the modern value of   $\rm d_{\small M31}=784$ kpc.  Top-left panel shows galaxies without solar motion correction, top-right panel corrected by Hubble's 1929 solar motion 
and bottom panel with the calculated solar motion correction. The $\rm H_\circ$ parameters obtained 
in all cases are near the accepted value nowadays of $\rm H_\circ=72\pm 10\%$ (km/s)/Mpc.    }
\label{fig:nearby}
\end{figure}

Figure \ref{fig:all} shows three diagrams: the sample without correction and a fit v=Kr, the sample 
corrected by Hubble's 1929 solar motion and a fit v=Kr, and the sample corrected for the 
solar motion, by the method of least squares applied to eq. \ref{eq:e4} and the resulting line v=Kr.
As we can see, Hubble's constant in all three cases is not substantially different. The obtained values 
are, however, smaller than the modern value of H$_\circ$=72 (km/s)/Mpc. We should 
compare, on the other hand, this value with the one found by Hubble (1929), H$_\circ$=465$\pm$50  
(km/s)/Mpc. Two factors contributed for the better performance of Lundmark's data: his method of 
distances, simpler and more reliable than Hubble's, and the use of the distance to M31, 
which was, of course, unknown at the time.   
The correction of the solar motion, as laid down by Hubble, varies with the depth (distance) of the 
galaxy sample. The largest distances are, because of the largest difficulty of observation, the most 
affected by uncertainties.  Accordingly, it would be interesting selecting a nearby 
Lundmark's subsample, whose distances would be better evaluated and, besides, 
similar to Hubble's 1929 distances.  Such a procedure would imply in an opportunity of 
a direct comparison with the result obtained by Hubble. We 
know now that Hubble underestimated  his distances by a factor of $\approx$10. Hubble's largest 
distance is 2.0 Mpc (see his Table 1). We can restrict Lundmark's sample to galaxies closer than 
10$\times$2.0=20 Mpc. Doing that results in a sample of 12 galaxies. We redid the procedure of 
Fig. \ref{fig:all}  and the result is in Fig. \ref{fig:nearby}.
Again, as we can see in Fig. \ref{fig:nearby}, the solar motion correction does not significantly affect 
the determination of $\rm H_\circ$. Incidentally, it is interesting to point out that the original sample 
of Hubble (excluding the negative velocities) {\it without solar motion correction} gives a linear 
correlation v=Kr, with K=446 (km/s)/Mpc, consistent with the value determined by him {\it after 
solar motion correction} (K = 465$\pm 50$). 

Qualitatively, the solar motion obtained with Lundmark's reduced sample is compatible with the one 
obtained by Hubble. The solar motion apex in Hubble ($\alpha=$19 hours and $\delta=$+40 degrees) 
sits approximately in the direction of the star Vega, the brightest star of the Lyra constellation. With 
Lundmark's 12-galaxy sample, the apex sits nearby, in the boundary between Lyra and Vulpecula 
constellations ($\alpha=$19 hours and $\delta=$+23 degrees). However, the uncertainty of distances 
in both samples makes almost irrelevant such similarities. 

\section{Final remarks}
In the modern relativistic cosmology, redshifts of distant objects are interpreted as the result of 
expanding space. For small values of z, as those discussed here, i.e., z$\ll$1, both 
interpretations --- Doppler effect and expanding space --- are mathematically indistinguishable 
(see Soares 2009, Fig. 4). The influence of the solar motion on galaxy velocities is only 
physically meaningful within the interpretation of redshifts as originating from the Doppler effect.  

The idea that there might be a component of systematic motion in the nebula velocities, as mentioned 
in the beginning of section 2, was not Hubble's. It had been introduced by the German astronomer 
Carl Wilhelm Wirtz (1876-1939) in 1918, following what was already done in the determination of 
the solar motion with respect to the stars (Hubble 1936, p. 107). Wirtz assumed then the existence 
of a {\it K-term} (or {\it K-correction}) as a constant velocity (K from the German {\it konstant}), which 
might be subtracted from the nebula velocities, before the solar motion determination: 
\begin{equation}
\label{eq:e6} 
v = K + X \cos \alpha\cos \delta + Y {\rm sen}~  \alpha\cos \delta + Z {\rm sen}~ \delta ~, 
\end{equation}
where K is the velocity  correction that should be applied to v. With such a correction and the 
removal of the solar motion, the situation of the velocity residuals of Wirtz' nebulae improved, but 
still was not entirely satisfactory: they did not distribute in a completely random way --- as expected 
--- and, in addition to that, the derived K-term was of about  800 km/s, intriguingly large and 
comparable to the resulting solar motion ($\approx$700 km/s). 

The situation would considerably improve with the introduction of the K-correction varying with 
distance, as done by Hubble in 1929 (cf. eq. \ref{eq:e4}). The solar motion determined by him, in 
such a way, was of about 300 km/s, in the approximate direction of Vega (Hubble 1936, p. 114), 
and the velocity residuals were satisfactorily random. 
Hubble's ingenuity  was his decision in adopting the simplest hypothesis for the variable K-correction, 
namely, of the type Kr, while other astronomers got lost in much more complicated --- 
and at that point unnecessary and even unjustifiable --- K(r) expressions. On the other hand, Hubble  
was  aware that the relationship v=Kr was consistent with the prediction of de Sitter's  cosmological 
model (Hubble 1929, p. 173). 

As we have seen, solar motion correction in the early extragalactic astrophysics did not turn out 
to be important in the determination of the theoretical expansion parameter, mainly because 
of the significant errors in distance determination. In modern extragalactic astrophysics and 
cosmology, however, distances and spectral shifts are determined with much better precision and 
solar motion correction becomes a fundamental aspect of the evaluation of theoretical parameters.  

Nowadays, the observations of galaxies outside the Local Group, from a given observatory, are 
submitted  to two corrections. First, as before, the heliocentric correction is done, and, in the second 
place, differently of what has been done above, the solar motion correction is done with respect to 
the barycenter --- or centroid --- of the Local Group of galaxies.  Velocities become then referred to 
to the center of the Local Group, and may then be used in the investigations of extragalactic 
issues, such as the expanding universe problem. For the technical details of these corrections 
see, for example, the articles by Yahil, Tammann and Sandage (1977) and by Karachentsev and 
Makarov (1996). 

\bigskip 

{\noindent\bf Acknowledgment -- }One of us (DS) would like to thank Alexandre Bagdonas Henrique for 
helpful discussion on Lundmark's work.

\end{document}